\documentclass[12pt, a4paper]{article}
\pdfoutput=1

\usepackage{amsmath}
\usepackage{amsfonts}
\usepackage{amssymb}
\usepackage{graphicx, rotating}
\usepackage{epsfig}
\usepackage{latexsym}
\usepackage{graphicx}
\usepackage{color}
\usepackage{amsmath,bm,amssymb}
\usepackage{cite}
\usepackage{slashed}
\usepackage{hyperref}
\hypersetup{colorlinks, citecolor=bluscuro, linkcolor=black, urlcolor=bluscuro}
\definecolor{rossos}{cmyk}{0,1,1,0.55}
\definecolor{bluscuro}{rgb}{0.15, 0.2, .85}
\definecolor{bluchiaro}{cmyk}{1,.3,0.,0.1}

\newcommand{\lp}{\left(}
\newcommand{\rp}{\right)}

\newcommand{\be}{\begin{equation}}
\newcommand{\ee}{\end{equation}}
\newcommand{\bea}{\begin{eqnarray}}
\newcommand{\eea}{\end{eqnarray}}

\newcommand{\arXiv}[2]{\href{http://arxiv.org/pdf/#1}{{\tt [#2/#1]}}}
\newcommand{\arXivold}[1]{\href{http://arxiv.org/pdf/#1}{{\tt [#1]}}}

\def\bma#1{\mbox{\boldmath{$#1$}}}

\def\vecphi{\bm{\phi}}

\begin{document}
\allowdisplaybreaks
\begin{titlepage}
\begin{flushright}
CERN-TH-2018-245\\
DESY 18-206
\end{flushright}
\vspace{.3in}

\vspace{1cm}
\begin{center}
{\Large\bf\color{black} 
A Fresh Look at the Calculation 
of Tunneling Actions \\[0.3cm] 
in Multi-Field Potentials} \\
\vspace{1cm}{
{\large J.R.~Espinosa$^{a,b,c}$, T.~Konstandin$^d$}
\vspace{0.3cm}
} \\[7mm]
{\it {$^a$\, Institut de F\'isica d'Altes Energies (IFAE), The Barcelona Institute of Science and Technology (BIST), Campus UAB, 08193 Bellaterra (Barcelona), Spain}}\\
{\it $^b$ {ICREA,  Instituci\'o Catalana de Recerca i Estudis Avan\c{c}ats,\\ Pg. Llu\'is Companys 23, 08010 Barcelona, Spain}}\\
{\it $^c$ {Theoretical Physics Department, CERN, CH-1211 Geneva 23, Switzerland}}\\
{\it $^d$ {DESY, Notkestr. 85, 22607 Hamburg, Germany}}
\end{center}
\bigskip

\vspace{.4cm}

\begin{abstract}
The quantum decay of a metastable vacuum 
is exponentially suppressed by a tunneling action that can be calculated in the semi-classical approximation as the Euclidean 
action of a bounce that interpolates between the false and true phases. For multi-field potentials, finding the bounce is non-trivial due to its peculiar boundary conditions and the fact that the action at the bounce is not a minimum but merely a saddle point.  Recently, an alternative tunneling action has been proposed that does not rely on Euclidean bounces and reproduces the standard result at its  minimum. Here we generalize this new approach for several scalar fields and demonstrate how its use can significantly improve the numerical calculation of tunneling actions for multi-field potentials.  
\end{abstract}
\bigskip

\end{titlepage}

\section{Introduction \label{sec:intro}} 

The phenomenon of tunneling decay of a metastable state via thermal or quantum fluctuations is ubiquitous in particle physics, condensed matter systems and cosmology.
In the semi-classical approximation, the decay probability is exponentially suppressed by the tunneling action in WKB approximation. The Euclidean approach~\cite{Coleman}
calculates this action via the so-called bounce solution that dominates the path integral in this regime. Finding the bounce solution in multi-field potentials is a non-trivial problem. The bounce starts from some {\em a priori} unknown release point and asymptotes to the metastable minimum of the potential at late times. To obtain the bounce numerically from the action is hindered by the fact that 
the bounce is not a minimum of the action but a saddle point. This is no accident and has physical meaning:
the second variational derivative of the Euclidean action at the bounce  is expected to have one (and only one) negative eigenvalue 
as a signal of the instability of the local minimum \cite{negative}. 

In the case of a single scalar field, the bounce solution can be easily found via the over-/under-shooting method \cite{Coleman} although the task is harder when false and true vacua are nearly degenerate (the so-called thin-wall regime. In that case one can resort to the semi-analytic thin-wall approximation \cite{Coleman} but its accuracy degrades quickly when the potential difference between vacua grows \cite{SH}). 
For several scalar fields, over-/under-shooting is typically not an option. With just two fields a combination of shooting and aiming can be successful but this ceases to be viable with more fields. 

In the literature, several approaches and algorithms can be found that aim at finding multi-field bounce solutions 
\cite{BBW,CHH,Kusenko,Dasgupta,MQSJ,CMS,KH,JL,Park,ALP,MOSW,Bounds}. 
In~\cite{Kusenko}, the action is modified by adding extra terms that lift the negative mode [but vanish on solutions of the Euclidean equations of motion (EoM)] thus facilitating the search for the bounce. In any case, the improved action has only a local minimum at the bounce and one needs a starting field configuration quite close to the bounce in order to converge to it.
 
In~\cite{CMS} an algorithm based on a combination of 'shooting' and 'cooling' was proposed.
In the shooting phase, the bounce is determined for a fixed path, while in the cooling phase, the path is changed using a gradient descent algorithm. The idea behind this construction is that the shooting part of the algorithm is mostly sensitive to the negative eigenvalue while the cooling is only sensitive to the positive eigenvalues.
This algorithm works in general but cooling and shooting partially compete with each other due to the negative eigenvalue also affecting the cooling, and convergence is rather slow. 

Alternatively, in~\cite{KH} a method was proposed that modifies the friction term in the bounce equation of motion with a factor $\alpha$. The procedure starts with a system without friction ($\alpha=0$) that can be easily solved by energy conservation. The system then slowly transitions to the original system ($\alpha=d-1$, where $d$ is the number of space-time dimensions), solving the EoM in every step using the Newton method. Since the Newton method converges to the closest critical point, the bounce is recovered. Still, also in this case the transition has to be performed very slowly in order to ensure convergence.

\begin{figure}[t!]
\begin{center}
\includegraphics[width=0.8\textwidth]{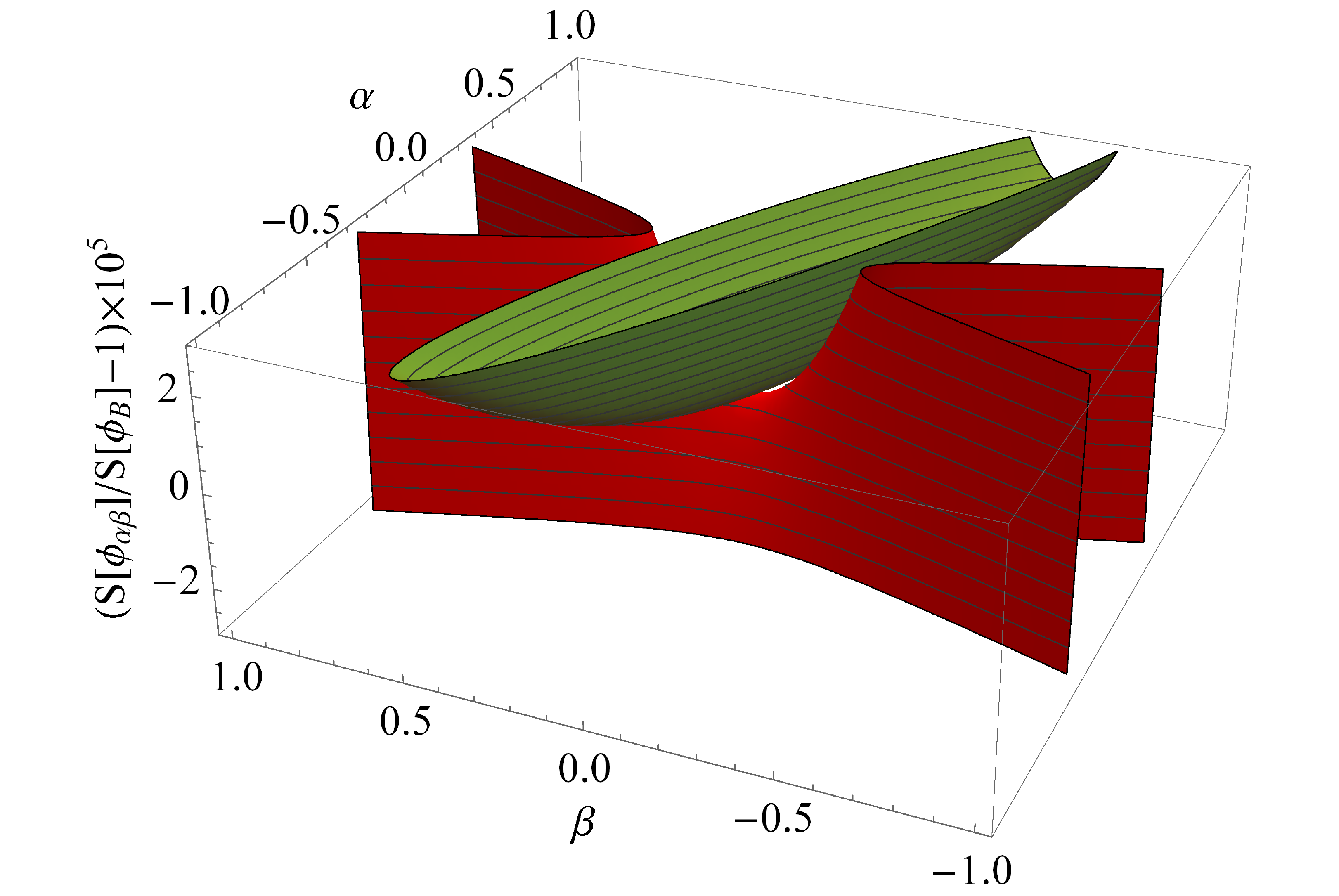}
\end{center}
\caption{\label{fig:Actions}  For $V(\phi)=-\lambda \phi^4/4$, the plot shows the Euclidean (red) and tunneling-potential (green)  actions around the bounce, given by the Fubini instanton $\phi_B(r)=\phi_0/(1+\lambda\phi_0^2r^2/8)$. The deformation around the bounce is parametrized as $\phi_{\alpha,\beta}(r)=\phi_B(r)+\alpha\, \phi_\alpha(r)+\beta\, \phi_\beta(r)$, where
$\phi_\alpha(r)=\phi_B(r)\sqrt{\lambda/(4\pi)}$ is the negative mode and  $\phi_\beta(r)=\sqrt{2/\pi}R(r^4+R^4-3r^2R^2)/(r^2+R^2)^3$, with $R^2=8/(\lambda\phi_0^2)$ (see {\it e.g.} \cite{lphi4}) is a positive mode. }
\end{figure}

Recently, it was noticed in~\cite{E} that the tunneling action for one field can be obtained from minimizing an alternative action. The new action is  written directly in field space in terms of a {\em tunneling potential} and the original time/space Euclidean coordinates (and the negative mode related to their rescaling) are removed from the problem.  To illustrate this point, Fig.~\ref{fig:Actions}, compares the behaviour of the standard Euclidean action (red) and the tunneling-potential action of Ref.~\cite{E} (green)
when deforming the bounce along two directions in field-configuration space. The direction parametrized by $\alpha$ corresponds to deformations of the bounce size. Along this direction   the action is a maximum at the bounce (this is the only negative mode). Instead $\beta$ corresponds to an orthogonal deformation with positive eigenvalue (for technical details, see caption).  These two directions illustrate that the Euclidean action has a saddle point at the bounce. The same deformations, translated to  the formulation in terms of the tunneling potential \cite{E}, lead to the behavior shown by the tunneling potential action (green), which has a minimum at the bounce. This nice property was already used in \cite{E} in the numerical evaluation of tunneling actions in the single-field case.

Since the second functional derivative of the Euclidean action at the bounce only contains one negative eigenvalue also in the multi-field case, it is expected that the tunneling potential formulation extended to many fields will enjoy the same appealing property of the single-field case discussed above. This would bypass the 
numerical issues with finding a saddle point and make it possible to determine the tunneling action with much more robust numerical recipes.  In this paper we present (Section 2) such a multi-field action, generalizing to an arbitrary number of fields  the tunneling-potential action of \cite{E}. In Section 3 we show that this action is an extremal under changes of the tunneling potential or the tunneling trajectory in field space. In Section 4 we present the algorithm used to find numerically the action illustrating it with examples of two-field potentials. In Section 5 we present our conclusions and outlook.

\section{Multi-field Tunneling Actions \label{sec:bounce}} 

To study tunneling in multi-field potentials, let us collect the scalar fields of the theory in question into ${\vecphi}$, a vector with components $\phi_n$ with $n=1 \dots N$ for $N$ fields (we used boldface for vectors in field space). The multi-field potential $V(\vecphi)$ has some false vacuum at $\vecphi_+$ and we are interested in studying the exponential suppression of the decay out of that false vacuum by quantum fluctuations. Without loss of generality we can take
$V(\vecphi_+)=0$ (in the absence of gravity) and $\vecphi_+={\bf 0}$.

In the Euclidean formulation \cite{Coleman}, the action that 
suppresses exponentially the decay rate of a false vacuum  is the Euclidean action ($a=1,\dots,4$ labels  the 4-dimensional Euclidean space coordinates $\{\tau,x_i\}$)
\be
S_E = \int d^4 x \left[\frac12 \partial_a\vecphi\cdot\partial_a\vecphi +V(\vecphi)\right]\ ,
\ee
evaluated on the bounce  solution $\vecphi_B$, a solution of the 
Euclidean  EoM that connects the false vacuum at $\vecphi_+$, with 
\be
\vecphi_B(\tau\rightarrow \pm\infty)=\vecphi_+\ , 
\label{BC1}
\ee
with some field configuration on the basin of the true vacuum, reached at $\tau=0$, with 
\be
\partial_a\vecphi_B(\tau=0)={\bf 0}\ .
\label{BC2}
\ee
For a single field ($N=1$) the bounce is $O(4)$ symmetric ({\it i.e.} a function of $r^2=\tau^2+x_ix_i$), as proved in Ref.~\cite{CGM}, and this leads to a dramatic simplification of the problem. For more than one field ($N>1$)
there has been recent significant progress towards proving  that 
the bounce is also $O(4)$ symmetric in this case \cite{Kfir}. Assuming then an $O(4)$ symmetric bounce
the Euclidean EoM takes the simple form (a dot representing a derivative with respect to $r$)
\be
\label{eq:eom}
\ddot{\vecphi} +\frac{3}{r}\dot{\vecphi} = \bm{\nabla} V\ ,
\ee
where $\bm{\nabla} V$, the field-space gradient of $V$, has components $\partial V/\partial\phi_n$.
The boundary conditions (\ref{BC1}) and (\ref{BC2}) take the form
\be
\dot{\vecphi}_B(0)=\bm{0}\ ,\quad \vecphi_B(\infty)=\vecphi_+\ .
\ee
Identifying $r$ with time, Eq.~(\ref{eq:eom}) corresponds to the classical motion of a particle in the inverted multi-field potential $-V(\vecphi)$ with a velocity and time-dependent friction force. Finding the solution requires scanning the value of the fields at the center of the Euclidean bubble, $\vecphi_B(r=0)\equiv\vecphi_0$, until the boundary condition at $r\rightarrow \infty$ is satisfied. This point $\vecphi_0$ is called the release (or escape) point. 
The Euclidean action on the bounce reads then (dropping from now on the subindex $B$ for simplicity)
\be
S_E[\vecphi] = 2\pi^2\int_0^\infty  \left[\frac12 |\dot{\vecphi}|^2
+ V(\vecphi)\right]r^3dr\ .
\label{SE}
\ee

In the following it is helpful to introduce $\varphi$ as canonical field along the tunneling path, with 
\be
d\varphi^2= d\vecphi \cdot d\vecphi\ .
\label{varphi}
\ee
That is, $\varphi$  is the arc length of the curve in field space. 
We have
\be
\bm{\nabla} V \cdot d\vecphi = \frac{\partial V}{\partial \phi_n}d\phi_n
\equiv\frac{d V}{d \varphi}d\varphi\equiv V'd\varphi\ ,
\label{gradV}
\ee
where a prime represents a $\varphi$ derivative and sum over repeated indices is implied. Notice that (\ref{varphi}) implies $\vecphi' \cdot \vecphi'=1$. A Frenet-Serret basis of orthonormal vectors can be introduced
along the bounce path as described in the Appendix.
The potential gradient can be decomposed in a term along the tangent to the path,  $\vecphi'$, and an orthogonal term as
\be
{\bma \nabla} V = V' \vecphi' + {\bma \nabla_{\perp} V}\ ,
\ee
with $\vecphi'\cdot {\bma \nabla_{\perp} V}=0$.

In general, the tunneling trajectory $\vecphi(r)$ is not straight in field space 
but the projection along the path reduces to the equation of motion of a one-field problem. Concretely, projecting (\ref{eq:eom}) onto $\vecphi'$ and noticing 
$\vecphi'' \cdot \vecphi'=0$, one finds
\be
\ddot{\varphi} +\frac{3}{r}\dot{\varphi} = V'\ .
\label{EoML}
\ee
The path curvature is determined by the projection of (\ref{eq:eom}) in a direction orthogonal to it. Such projection shows
that the curvature vector $\vecphi''$ is aligned with ${\bma \nabla_{\perp} V}$:
\be
 \dot\varphi^2 \vecphi''= {\bma  \nabla_{\perp} V}\ .
\label{EoMT}
\ee
We refer to Eq.~(\ref{EoML}) as the longitudinal (scalar) bounce equation and to Eq.~(\ref{EoMT}) as the transverse (vectorial) one.

The new approach to tunneling action calculations presented in \cite{E}
uses  an auxiliary function, $V_t(\varphi)$, the {\em tunneling potential}. In the language of the bounce discussed above, it is defined, for the multi-field case, as
\be
V_t(\varphi)\equiv V(\vecphi) -\frac12 |\dot{\vecphi}|^2\ ,
\label{Vt}
\ee
where $\vecphi$ and $\dot{\vecphi}$ in this expression should be understood as functions of the scalar field $\varphi$ (the canonical field that parametrizes the tunneling path).
 
Some properties of $V_t(\varphi)$ that are also true in the multi-field case are:

\noindent 1) $V_t(\varphi)$ is a monotonic function, with $V_t(\varphi)\leq V(\vecphi(\varphi))$. Notice 
that $V_t(\vecphi)$ is just minus the Euclidean energy. As the Euclidean energy is dissipated by the friction term in (\ref{eq:eom}):
\be
\frac{d}{dr}\left[\frac12 |\dot{\vecphi}|^2-V(\vecphi)\right]=-\frac{3}{r}|\dot{\vecphi}|^2\leq 0\ ,
\ee
it decreases monotonically as a function of $r$ \cite{CGM}. Since $\varphi$ is monotonous in $r$, monotonicity of $V_t(\varphi)$ follows.

\noindent 2) $V_t(\varphi)$ is in principle only defined along the tunneling path between $\vecphi_+$ and $\vecphi_0$. At these end points, $V_t=V$, as $\dot{\vecphi}_B(0)=\dot{\vecphi}_B(\infty)=0$.

We can follow the strategy of \cite{E} to remove altogether the reference to the bounce (and the 
4-dimensional Euclidean space in which it lives) in favor of $V_t(\varphi)$ also in the multi-field case. From (\ref{Vt}), using 
$\dot\vecphi\cdot\dot\vecphi=\dot\varphi^2$, we have
\be
\dot\varphi = - \sqrt{2[V(\varphi)-V_t(\varphi)]}\ ,
\label{dphi}
\ee
where the minus sign is chosen due to our convention $\varphi_+=0<\varphi_-$.  The Euclidean
radial coordinate $r$ can be extracted from (\ref{eq:eom})
as
\be
r=\frac{3|\dot{\vecphi}|^2}{\dot{\vecphi}\cdot (\bma\nabla V-\ddot{\vecphi_B})}=3\sqrt{\frac{2(V-V_t)}{(V_t')^2}}\ ,
\label{eq:r}
\ee
where $V_t'\equiv dV_t/d\varphi$.
Taking a derivative of the above equation with respect to $r$, we arrive at a differential equation for $V_t$:
\be
\left(4V_t'-3 V' \right)V_t' = 6(V_t-V)V_t'' \, ,
\label{VtEoM}
\ee
that takes the same form as the single-field case, but now $\varphi$
is the field along a curved path. This equation takes the place of (\ref{EoML}) in the new formulation. In addition, the transverse bounce equation (\ref{EoMT}) now reads
\be
2(V-V_t)\vecphi'' = \bma\nabla_\perp V\ .
\label{VtEoMT}
\ee

The tunneling action (\ref{SE}) can then be rewritten in terms of $V_t(\varphi)$ as
\be
S[V_t]=54 \pi^2\int_{\varphi_+}^{\varphi_0}\frac{(V-V_t)^2}{\left(-V_t'\right)^3} d\varphi \, .
\label{newSE}
\ee
We close this Section with some comments on this action.

Equation (\ref{newSE}) makes transparent the scaling of 
the action under changes of the shape of the potential, $V(\varphi)\rightarrow g^a V(\varphi/g^b)$, with $g>1$ some constant factor.
For $a>0$ ($a<0$) potential barriers increase (decrease) and for
$b>0$ ($b<0$) the width of the barriers increase (decrease). From Eq.~(\ref{VtEoM}), the scaling of $V_t$ is the same as for $V$,
so that 
\be
S\rightarrow g^{4b-a}S \ .
\ee
This results shows that, as expected, wider barriers suppress tunneling while it seems to imply that taller barriers facilitate it \cite{LargeN2}. 
This is due to the fact that $a>0$ increases the height of the barriers but also the depth of the potential minimum, which is crucial
for tunneling in QFT, contrary to the case in quantum mechanics. 
If one increases the height of the barrier leaving the true minimum
fixed, tunneling is also more suppressed. Indeed, for very large barriers
(for which the thin-wall approximation is applicable)
we can consistently take $(V-V_t)\rightarrow g^a(V-V_t)$ while keeping $V_t\sim V_t$ and then, $S\rightarrow g^{2a}S$.

For the case of a separable potential [like
$V(\phi_1,\phi_2)= V_1(\phi_1)+V_2(\phi_2)$ in the two-field case, with a false vacuum at $\vecphi_+={\bf 0}$ and $V(0,0)=0$], the original Euclidean action is clearly additive.
So, if we know bounce solutions $\phi_{iB}(r)$ for each potential separately, the path $\vecphi_B(r)=\{\phi_{1B}(r),\phi_{2B}(r)\}$
is also a solution of the  EoM for the two-field problem and its associated action is $S_E[\vecphi_B]=S_{E,1}[\phi_{1B}]+S_{E,2}[\phi_{2B}]$. On the other hand, the new action (\ref{newSE}) is clearly not additive in such cases. However there is no contradiction as the combined bounce $\vecphi_B(r)$ above is not the true bounce solution that controls the vacuum decay in this case. This is clear if one considers that around such bounce there are two negative modes, as each separate bounce $\phi_{B,i}(r)$ contributes one negative mode (related to the rescaling of its $r$ coordinate, leaving the other unchanged). The true action for tunneling out of the vacuum
at $\vecphi_+$ is  $S_E={\rm min}\{S_{E,1}[\phi_{B,1}],S_{E,2}[\phi_{B,2}]\}$  and the new action  (\ref{newSE}) is minimized at that same value.

\section{Extremality of the New Tunneling Action \label{sec:extr}} 

In going to the new formulation of the tunneling action, there is
freedom in the form of the action density in terms of $V_t$
(ultimately due to Derrick's theorem\footnote{Upon rescaling the 
bounce solution as $\phi_a(r)=\phi_B(r/a)$, one has $S_E[\phi_a]=a^2 S_K[\phi_B]+a^4 S_V[\phi_B]$, where $S_K$ ($S_V$) is the gradient (potential) part of the action. Stationarity of the action at the bounce, $dS_E[\phi_a]/da|_{a=1}=0$ leads to $S_K[\phi_B]/2=-S_V[\phi_B]=S_E[\phi_B]$, so we can write $S_E[\phi_B]=\alpha S_K[\phi_B]+(1-2\alpha)S_V[\phi_B]$ with arbitrary $\alpha$. Note also that $S_E[\phi_a]=(2-a^2)a^2S_E[\phi_B]$, explicitly showing that the action has a maximum (at $a=1$) along the rescaling deformation of the bounce.} \cite{Derrick}). It is important to choose the density in such a way that functional variation returns the right EoM. The aim of the current section is to show that requiring the action (\ref{newSE}) to be stationary under variations of the path and of $V_t$ returns the EoMs found in the previous Section [Eqs.~(\ref{VtEoM}) and (\ref{VtEoMT})]:
\bea
\label{VtL}
\left(4V_t'-3 V' \right)V_t' &=& 6(V_t-V)V_t'' \, ,\\
2(V-V_t)\vecphi'' &=& \bma\nabla_\perp V\ .
\label{VtT}\eea
These equations are equivalent to the usual Euclidean EoMs, either in its form (\ref{eq:eom}) or split into
the scalar EoM in (\ref{EoML}) and the vectorial one in (\ref{EoMT}). 

For this task it is useful to rewrite the action introducing an additional variable $\alpha$ that parametrizes the path. The action then reads 
\be
\label{eq:new_action}S=
54 \pi^2 \int_{\alpha_0}^{\alpha_+} \frac{(V-V_t)^2}{ \lp dV_t/d\alpha \rp^3} 
\lp \frac{d\vecphi}{d\alpha} \cdot \frac{d\vecphi}{d\alpha} \rp^2 d\alpha \, .
\ee
%
%
%

Notice that this action is invariant under a reparametrization $\alpha \to \bar \alpha(\alpha)$. A generic change of path $\vecphi$ and
tunneling potential  $V_t$ can be decomposed into a change of $V_t$ that keeps $\vecphi$ fixed and a change of path that keeps $V_t$ fixed. In the latter case, if the length of the path is modified, we can still keep $V_t$ fixed by using the $\alpha$ reparametrization invariance. Counting degrees of freedom, naively one might expect to get $N+1$ Euler-Lagrange equations from the stability of the action under $V_t$ and $\vecphi$ changes. However, a change $\delta\vecphi$ proportional to $\vecphi$ itself is equivalent to a change in $V_t$  so that one ends up with just $N$ independent Euler-Lagrange equations. In other words, the additional degree of
freedom corresponds to the $\alpha$ reparametrization invariance.

Let us first consider variations with respect to $V_t(\alpha)$ keeping $\vecphi$ fixed and setting $\alpha=\varphi$. Stability of the action leads to the Euler-Lagrange equation (here $s$ is the action density)
\be
\frac{d}{d\varphi}\left(\frac{\partial s}{\partial V'_t}\right)=\frac{\partial s}{\partial V_t}\ ,
\ee
leading to
\be
 \frac{3}{V_t^\prime} 
\left( 
-2 \frac{V_t^{\prime\prime}}{V_t^\prime} +  \frac{V^\prime - V_t^\prime}{V-V_t} 
+ 2 \frac{\vecphi^{\prime\prime}\cdot \vecphi^\prime}{\vecphi^{\prime}\cdot \vecphi^\prime}
\right)  = \frac{1}{V-V_t}  \, .
\ee
Noticing that $\vecphi^{\prime\prime}\cdot \vecphi^\prime = 0$
we find 
\be
-6 V_t^{\prime\prime} (V-V_t) -  (V_t^\prime)^2 + 3 V_t^\prime (V^\prime- V_t^\prime) = 0 \, ,
\ee
which is the EoM  in (\ref{VtL}).

Next, consider the stationarity of the action under variations of the path $\vecphi$.  We keep $V_t$ fixed with $V_t(\vecphi)=V_t(\vecphi+\delta\vecphi)$.\footnote{This cannot be done if 
$\vecphi_0$ is also varied in the modified path ($\vecphi_+$ is always fixed at the false vacuum) and $V(\vecphi_0+\delta\vecphi_0)\neq V(\vecphi_0)$. In that case we can always extend the original (or the deformed) path from $\vecphi_0$ to $\vecphi_0+\delta\vecphi_0$ keeping $V_t=V$ and $V_t'\neq 0$ in that interval. This extended path has the same action as the un-extended one and we can apply the argument below to it.}
If $\varphi$ is the canonical field along the original path, we can always choose
$\alpha=\varphi$ for the deformed path, even if it is not the
canonical field on that path. This choice ensures in addition $V_t'(\vecphi)=V_t'(\vecphi+\delta\vecphi)$.
Then the Euler-Lagrange equations
\be
\frac{d}{d\varphi}\left(\frac{\partial s}{\partial \phi_n'}\right)=\frac{\partial s}{\partial \phi_n}\ ,
\ee
give
\be
\label{eq:EoMphi}
2 \frac{d}{d\varphi} \left[ \vecphi^\prime
\frac{(V-V_t)^2}{V_t^{\prime 3}} \phi^{\prime 2}
\right] =\frac{(V-V_t)}{V_t^{\prime 3}} \phi^{\prime 4}\bm{\nabla} V \, ,
\ee
where we used $\phi^{\prime 2}\equiv \vecphi^\prime\cdot
\vecphi^\prime$. Using further $\phi^{\prime 2}=1$, we get
\be
2 \frac{d}{d\varphi} \left[ 
\frac{(V-V_t)^2}{V_t^{\prime 3}}\right]\vecphi^\prime
+2 \left[ 
\frac{(V-V_t)^2}{V_t^{\prime 3}}\right] \vecphi^{\prime\prime}
=\frac{(V-V_t)}{V_t^{\prime 3}} \bm{\nabla} V \, ,
\label{inter}
\ee
and projecting on the direction orthogonal to $d\vecphi$ the first term drops and we recover (\ref{VtT}). Projection of Eq.~(\ref{inter}) along the path gives again (\ref{VtL}): if $\delta\vecphi\propto d\vecphi$ one is not really changing the path but
$V_t$ instead, so one recovers the same equation obtained by varying $V_t$.

In essence, the problem formulated using the new action (\ref{eq:new_action})
is two-fold: find a path $\vecphi(\varphi)$ and a tunneling potential $V_t(\varphi)$ that minimize the action with the boundary conditions:
\be
V_t(\varphi_+)=V(\vecphi_+)\ ,\quad V_t(\varphi_0)=V(\vecphi_0)\ .
\label{BCV}
\ee
Notice that Eq.~(\ref{VtEoM}) then leads [using $V'(\vecphi_+)=0$] to
\be
V'_t(\vecphi_+)=0\ ,\quad V'_t(\vecphi_0)=3 V'(\vecphi_0)/4\ .
\label{BCVp}
\ee
In the next section we provide  numerical examples of this procedure.

\section{Numerical Examples\label{sec:example}} 

We present in this Section an efficient and simple algorithm  for numerical calculations of tunneling actions in multi-field potentials
that is based upon the new action of (\ref{newSE}) [or (\ref{eq:new_action})].
To illustrate the use of this algorithm we apply it to several examples with two-field potentials, selected to target some of the common difficulties encountered in such calculations.

The numerical algorithm we use is based on the following discretization of (\ref{eq:new_action}): 
\be
\label{eq:Sdisc}
S =\frac{27 \pi^2}{2}\sum_i [V(\vecphi_i)+V(\vecphi_{i+1})-V_{t,i}-V_{t,i+1}]^2 
 \frac{[(\vecphi_{i+1}-\vecphi_{i})\cdot(\vecphi_{i+1}-\vecphi_{i})]^2}{(V_{t,i+1}-V_{t,i})^3} \, .
\ee
For the path parameter $\alpha$ we use $V_t$ and we take $V_{t,i}$ as a fixed vector so that the action $S$ to be minimized is a function of the values of $\vecphi_i$. This can be done due to the monotonic nature of $V_t$ and  has the advantage that the action cannot become singular. More precisely, the vector $V_{t,i}$ is fixed for a given release point $\vecphi_0$ while we minimize the action with respect to the path $\vecphi_{i}$ and the release point $\vecphi_0$. In order to improve the numerical stability, the values $V_{t,i}$ should be more dense towards the beginning and end of the path. We use 
\be
V_{t,i} = V(\vecphi_+) +  \, x_i^2  \,(3 - 2 x_i)  [ V(\vecphi_0) - V(\vecphi_+) ] \, ,
\ee
with equidistant $x_i$ in $[0,1]$.
We use a Newton method that utilizes the first two derivatives of the discretized action (\ref{eq:Sdisc}) with respect to the path $\vecphi_{i}$ in order to minimize the action.

\begin{figure}[t!]
\begin{center}
\includegraphics[width=0.45\textwidth]{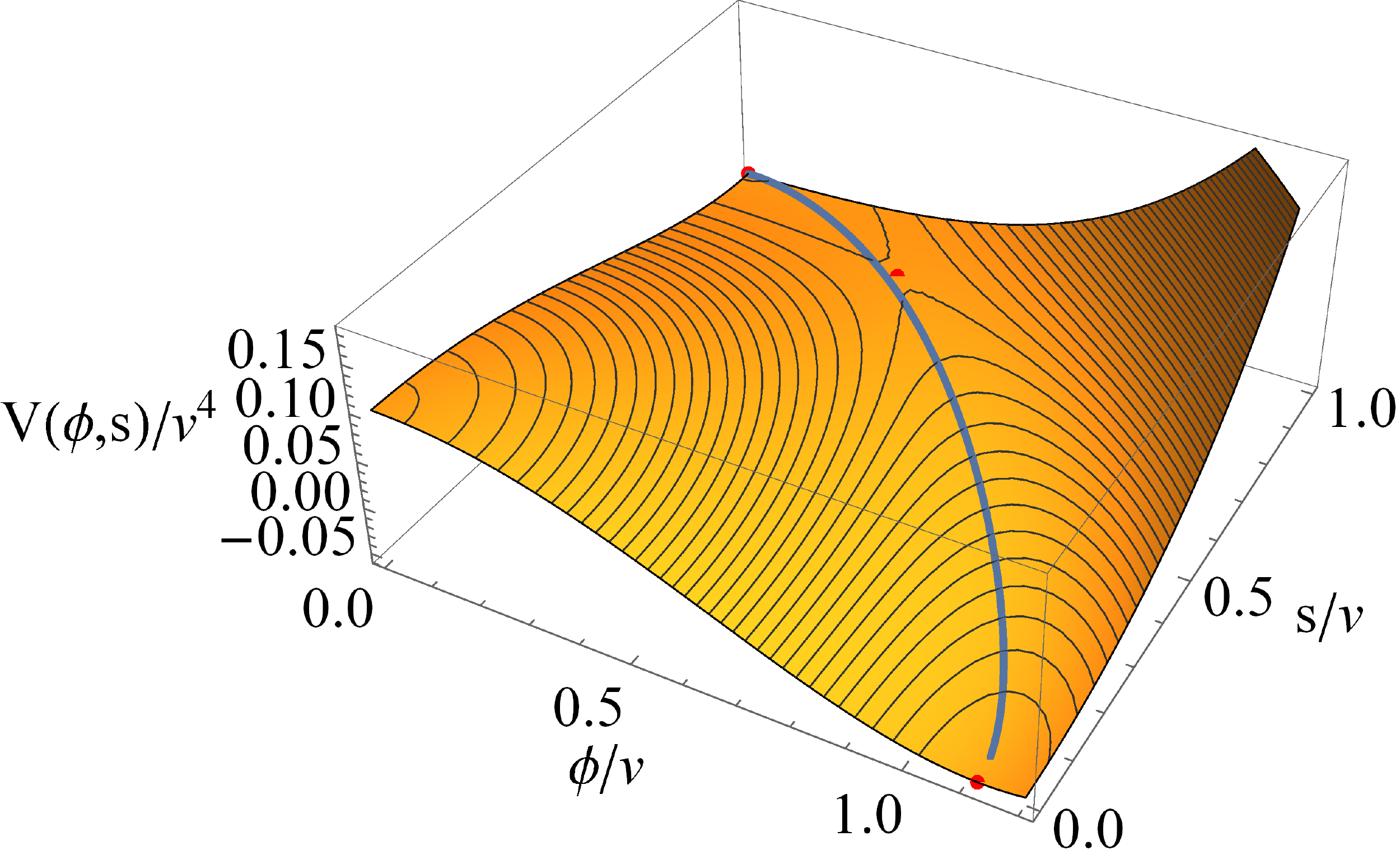}
\includegraphics[width=0.45\textwidth]{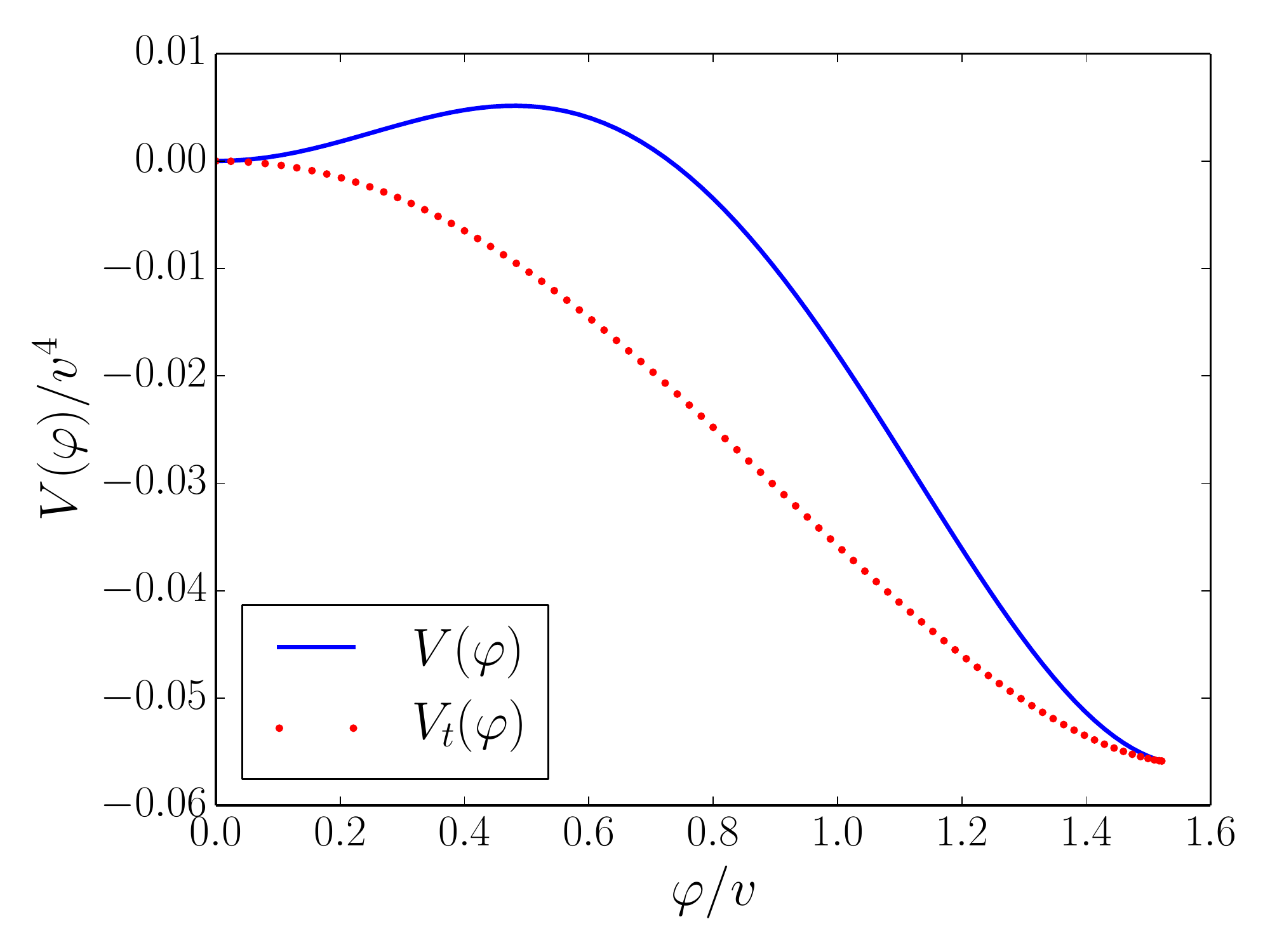}
\end{center}
\begin{center}
\includegraphics[width=0.45\textwidth]{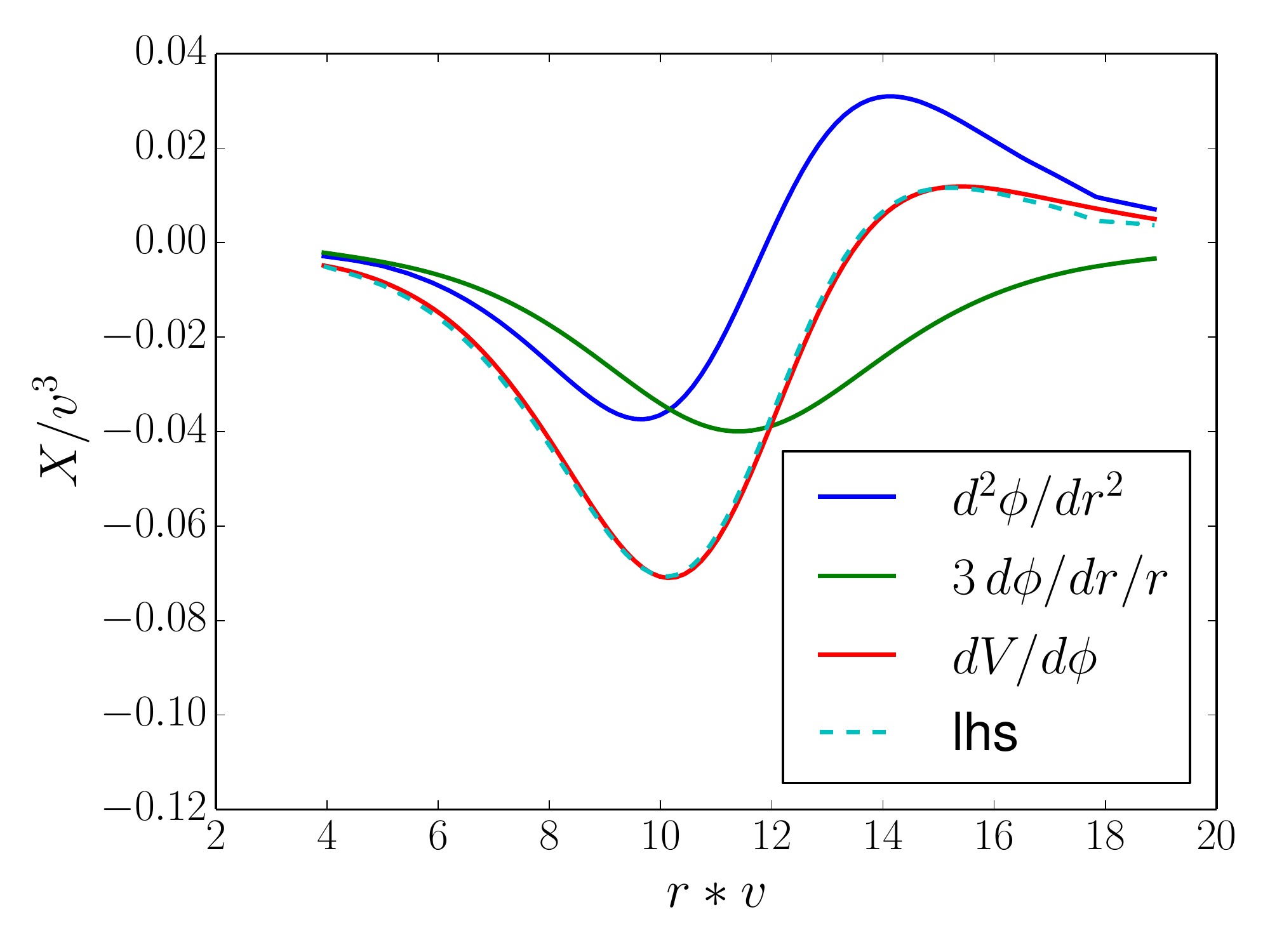}
\includegraphics[width=0.45\textwidth]{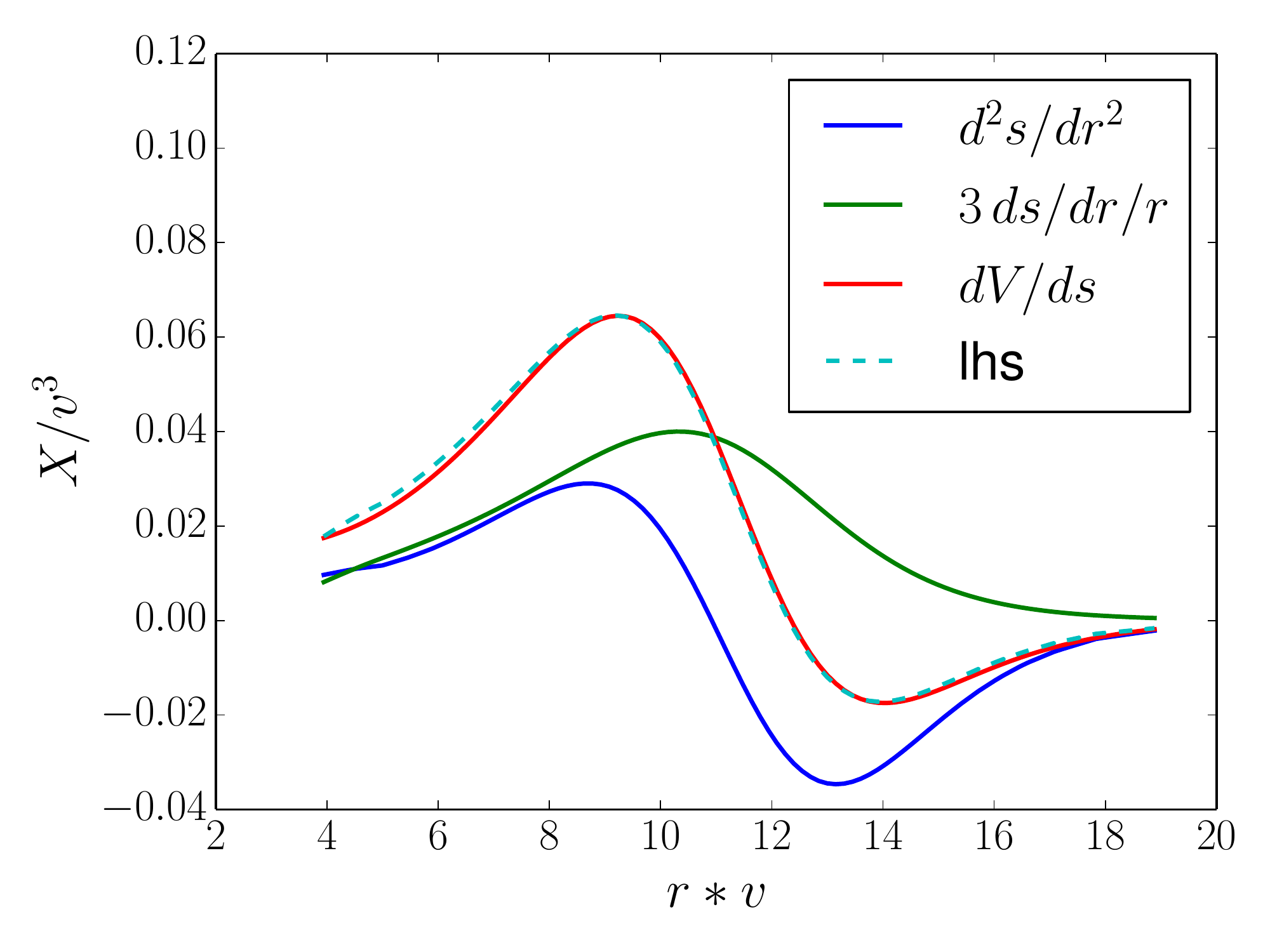}
\end{center}
\caption{\label{fig:A} 
For model A, Upper Left: trajectory in field space of the tunneling solution. Red dots mark the minima and saddle points in $V(\phi,s)$. Upper Right: potentials $V$ and $V_t$ along the tunneling path, as functions of the canonical field $\varphi$.
Lower Plots: Different contributions to the Euclidean EoM (\ref{eq:eom}) for $\phi$ (left) and $s$ (right). The EoM is  fulfilled at the percent level.
}
\end{figure}
\begin{figure}[t!]
\begin{center}
\includegraphics[width=0.45\textwidth]{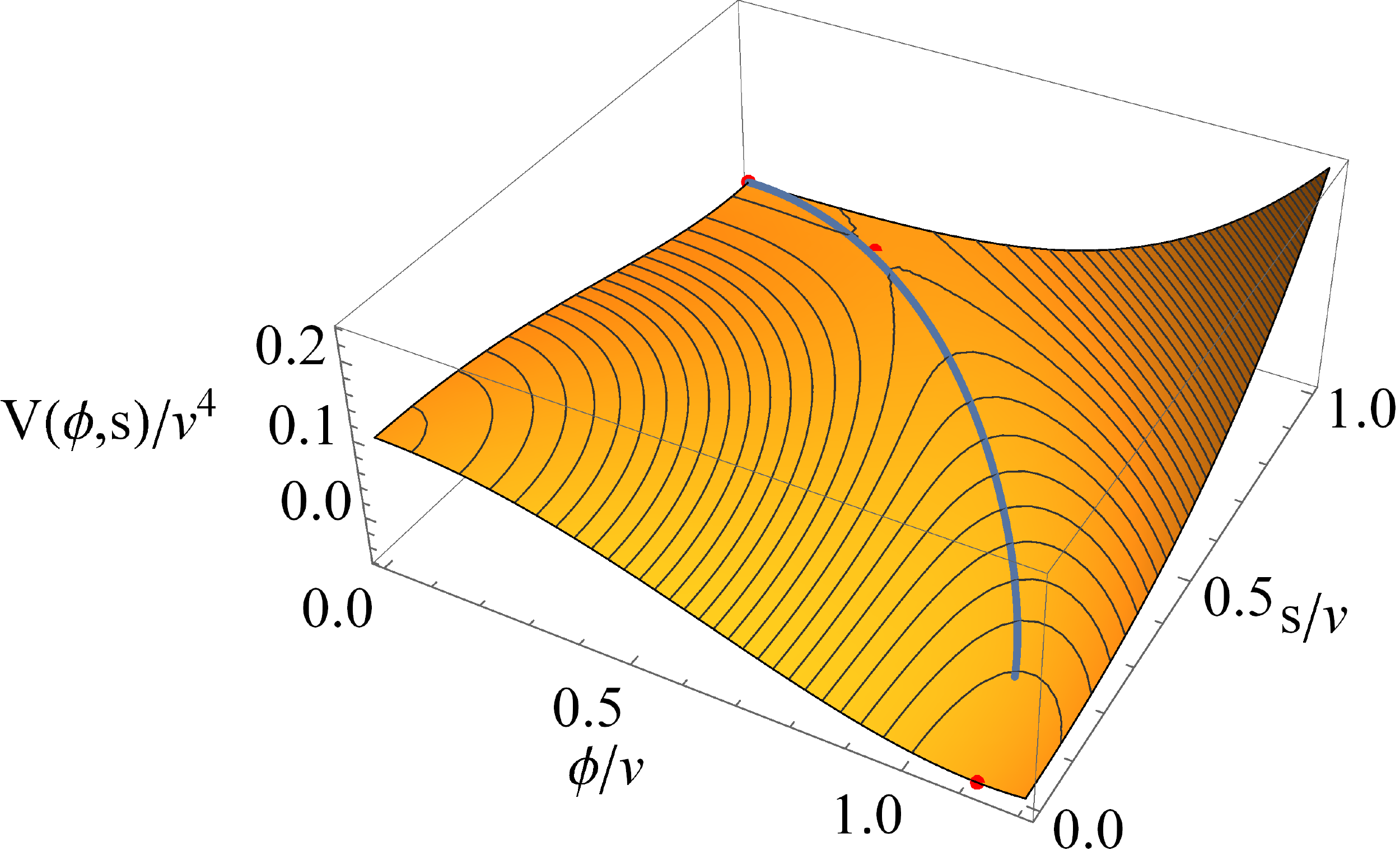}
\includegraphics[width=0.45\textwidth]{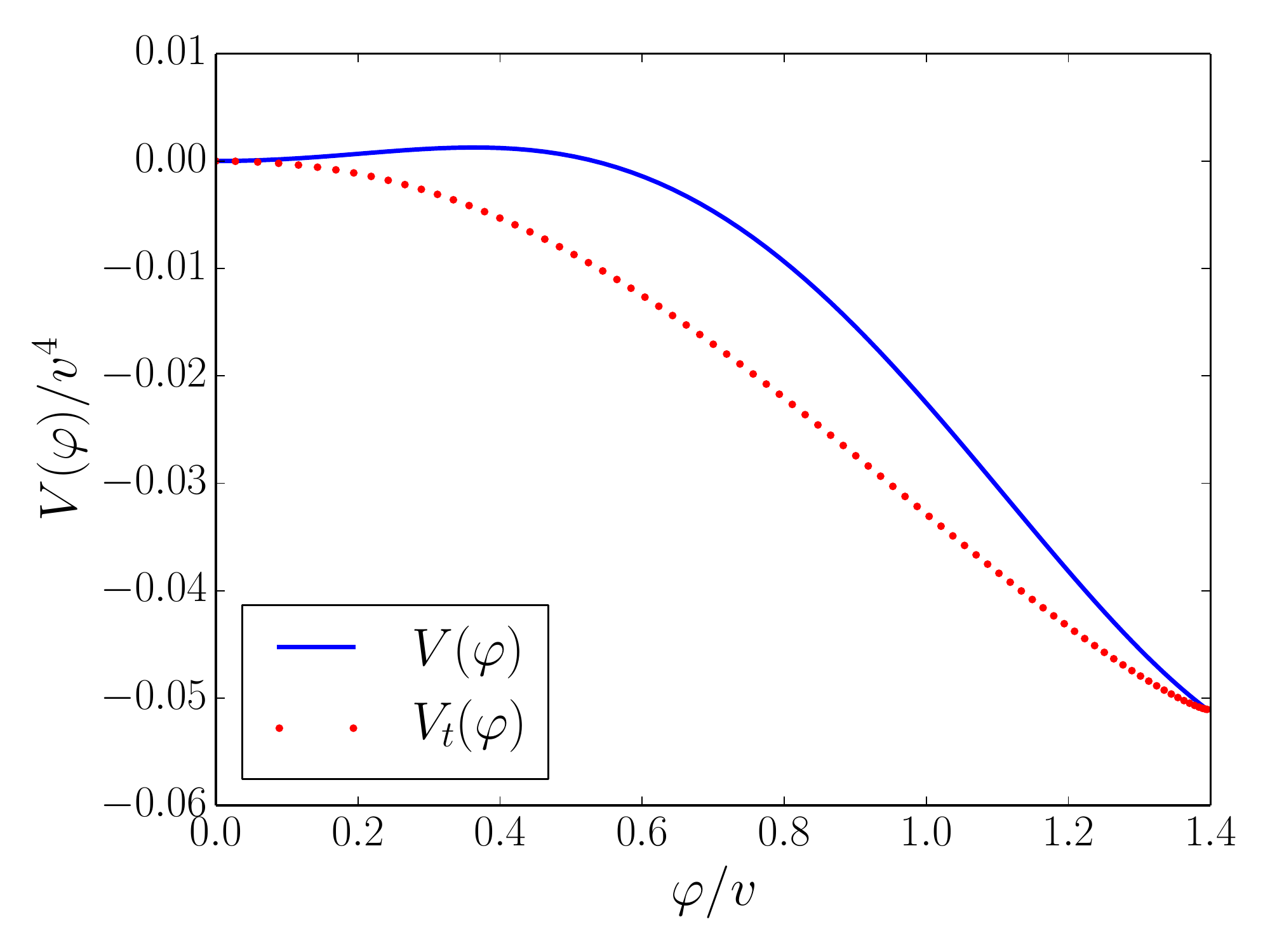}
\end{center}
\begin{center}
\includegraphics[width=0.45\textwidth]{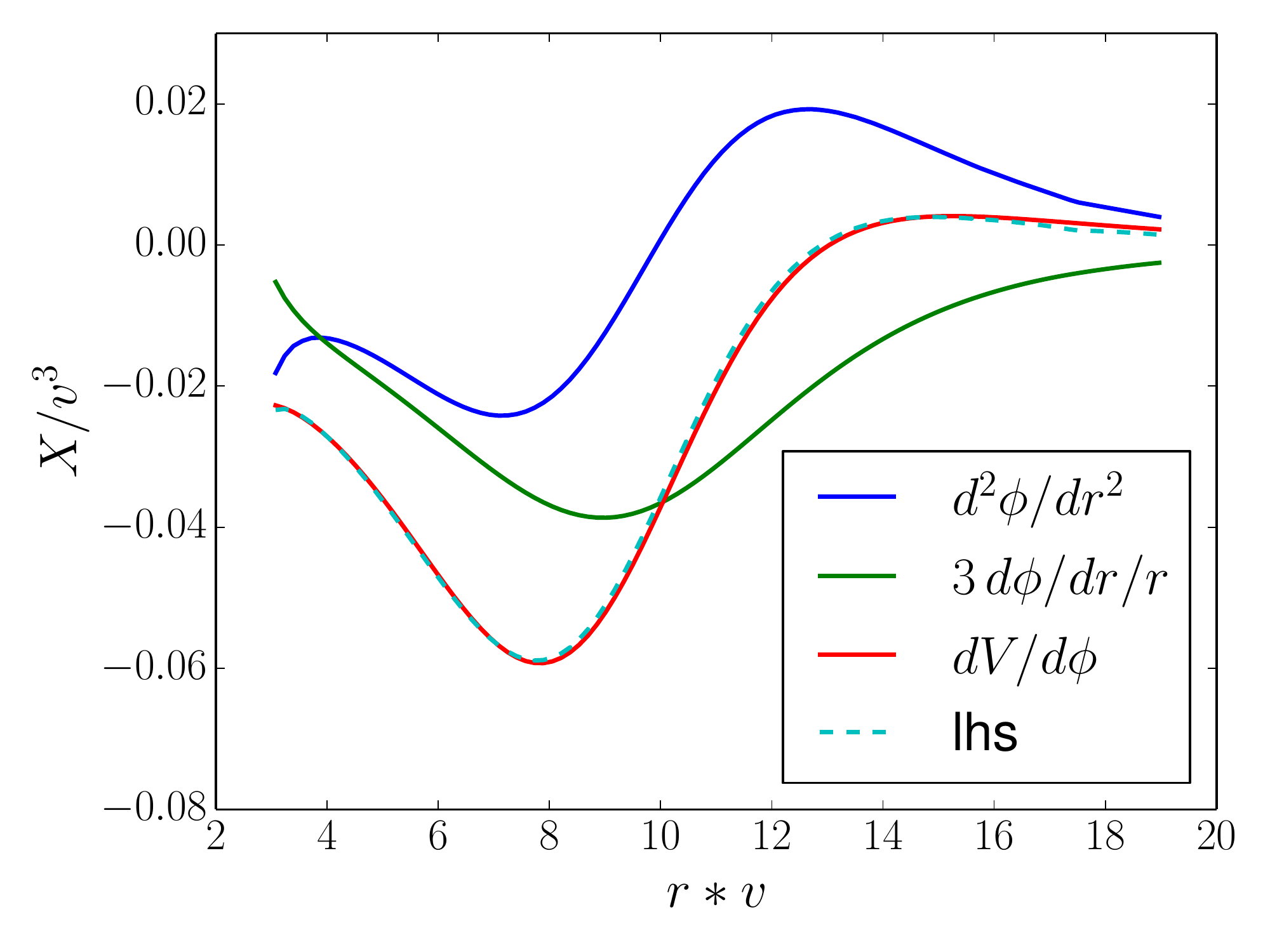}
\includegraphics[width=0.45\textwidth]{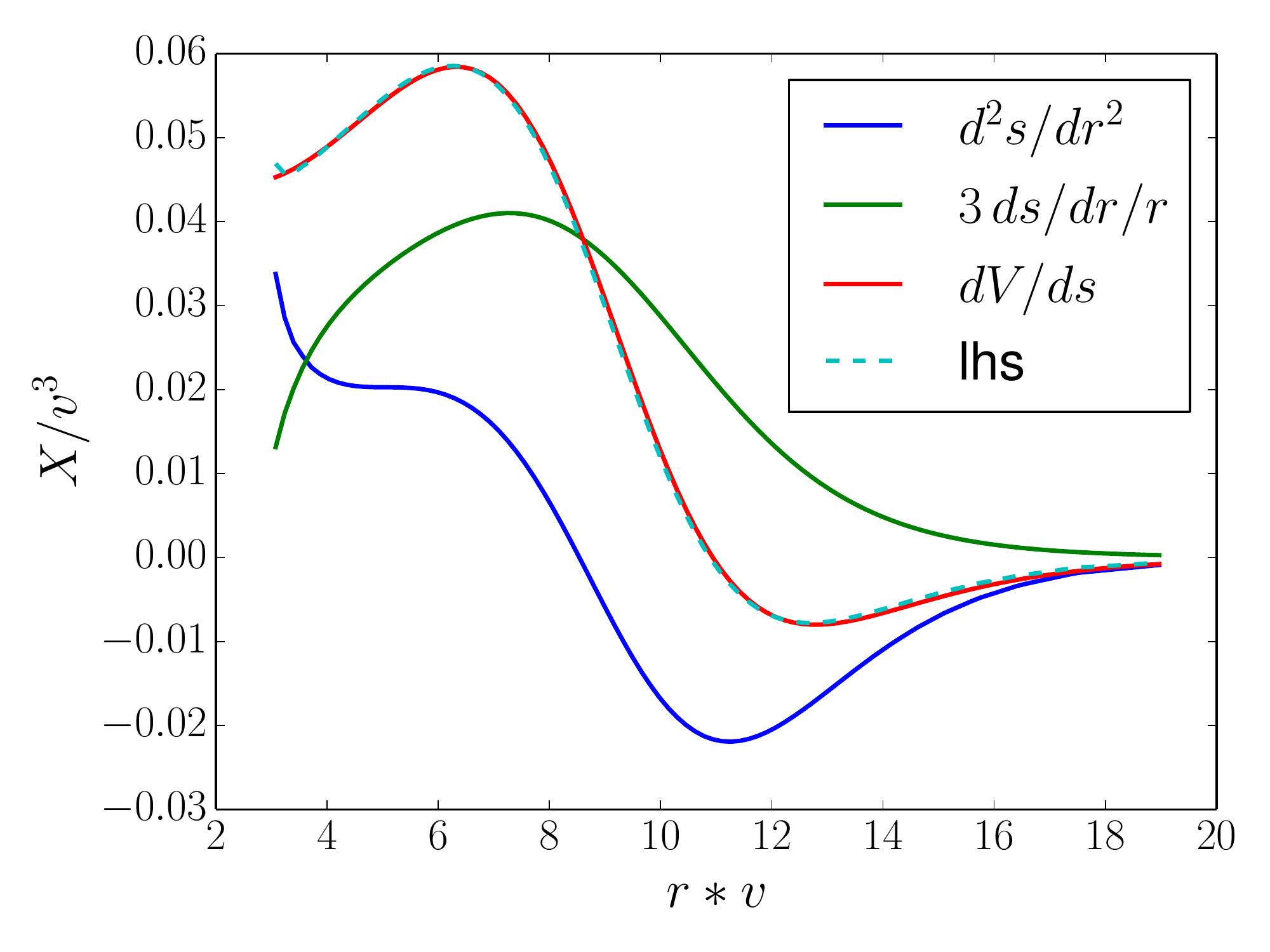}
\end{center}
\caption{\label{fig:A2}
Same as Fig.~\ref{fig:A} but for model A2.}
\end{figure}

Consider first the following potential with two scalar fields, $\phi$ and $s$,
\be
V(\phi,s) = \lambda (\phi^2 + s^2 - v^2)^2 + \lambda_b \phi^2 s^2 - \mu^2 \phi^2 \, ,\quad (\rm{model \, A}).
\ee
The model stems from a singlet extension of the Standard Model that is interesting for the two-step nature of the electroweak phase transition. The parameter $\lambda_b$ controls the height of the barrier between the local minima 
that break $Z_2$ at $(\phi= 0, s\not= 0)$ and the local minima that break the electroweak symmetry at $(\phi\not= 0, s= 0)$. The parameter $\mu$ lowers the electroweak vacuum such that it is the global minimum of the potential.
For the numerics we take first $\lambda = 0.1$, $\lambda_b = 0.1$ and $\mu^2 = 0.05 \, v^2$.
\begin{figure}[t!]
\begin{center}
\includegraphics[width=0.45\textwidth]{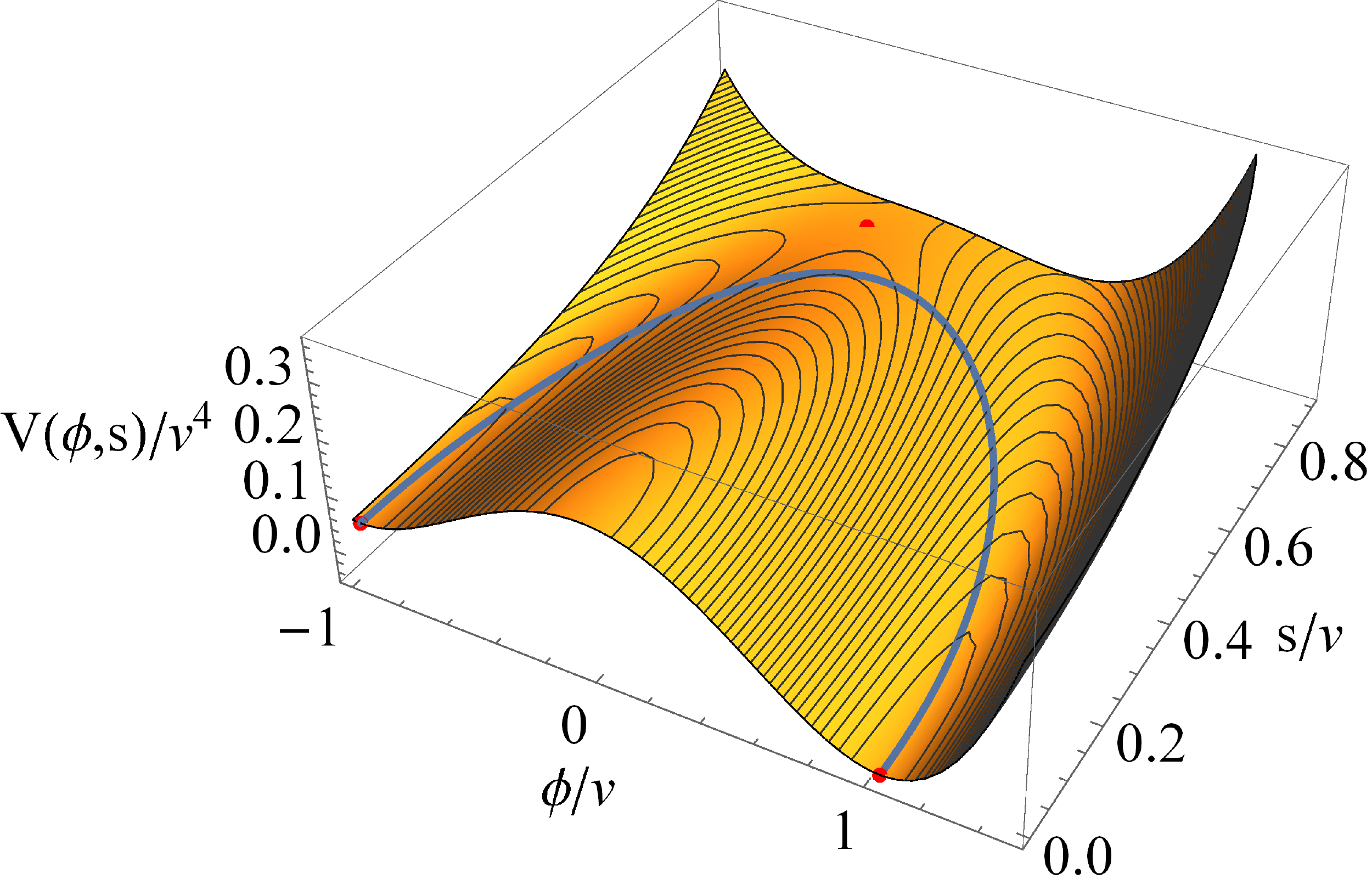}
\includegraphics[width=0.45\textwidth]{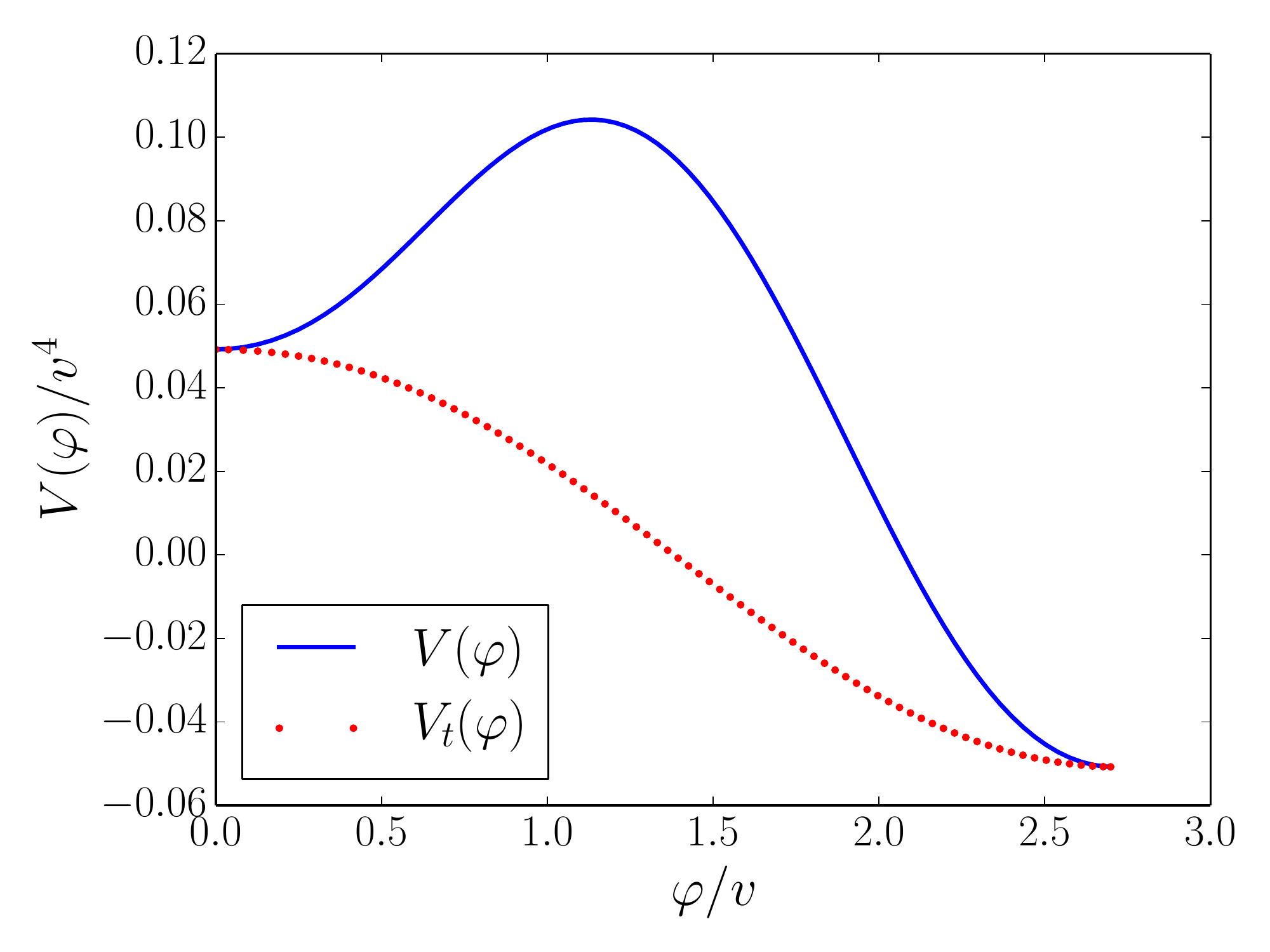}
\end{center}
\begin{center}
\includegraphics[width=0.45\textwidth]{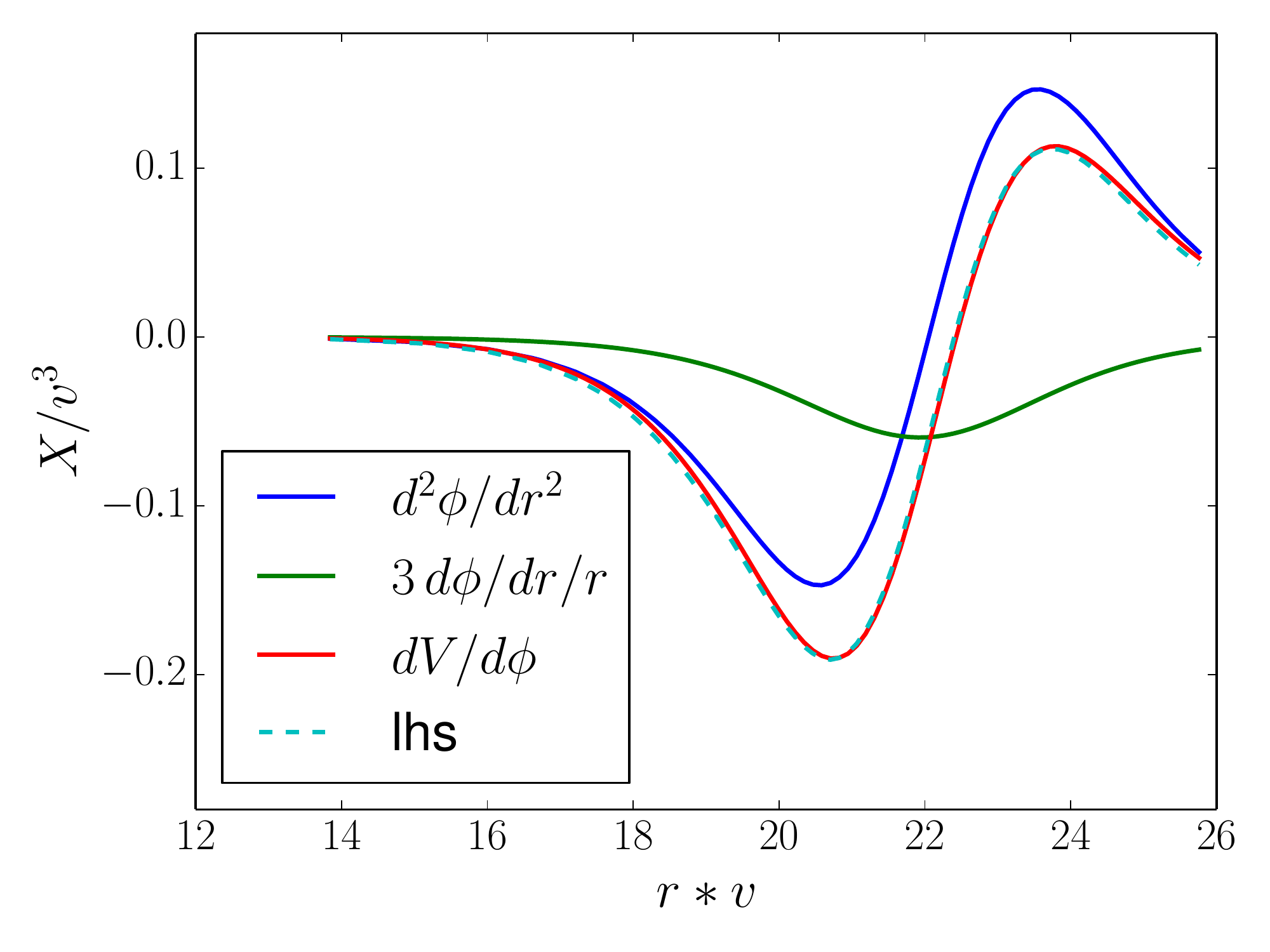}
\includegraphics[width=0.45\textwidth]{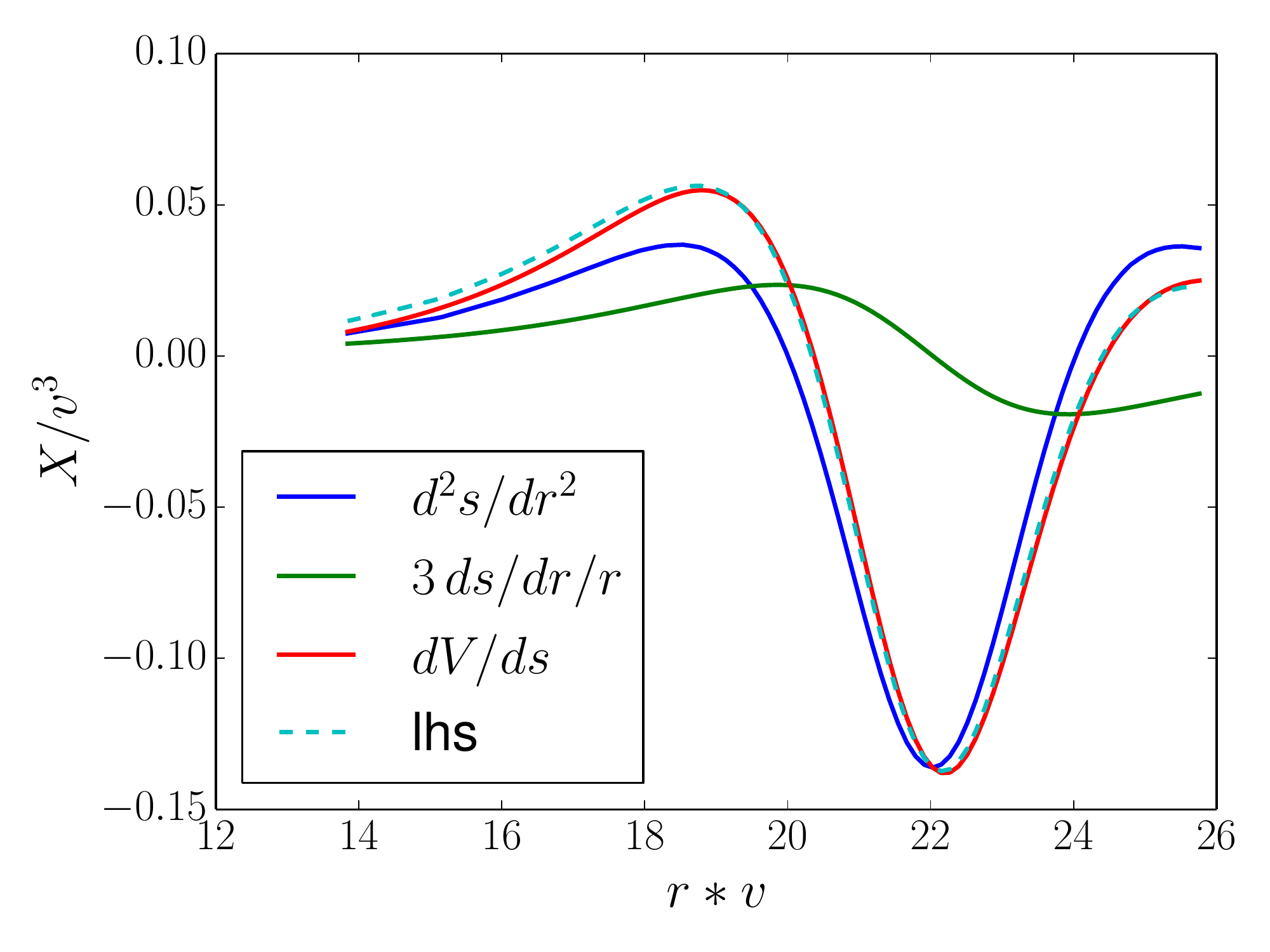}
\end{center}
\caption{\label{fig:B}
Same as Fig.~\ref{fig:A} but for model B. }
\end{figure}

Figure~\ref{fig:A}, upper left plot, shows the trajectory in field space that minimizes the discretized action (\ref{eq:Sdisc}) for 81 grid points on the path. As expected, the tunneling solution passes close to the saddle point of the potential but is somewhat repelled away from it in order to compensate the centrifugal forces. The action of the path is $S \simeq 1716$. The potential along the tunneling trajectory in terms of the canonical field $\varphi$, as well as the tunneling potential $V_t(\varphi)$ evaluated on the grid points, are shown on the upper right plot of Fig.~\ref{fig:A}.
As a crosscheck, we plot the different contributions to the Euclidean EoM in the lower plots of Fig.~\ref{fig:A}. The time coordinate $r$ is recovered via the relation (\ref{eq:r}) and the various derivatives are taken numerically from an interpolation of the path $\vecphi(r)$ through the 81 grid points. The check works reasonably well considering that the EoM involves a second derivative and that the number of grid points is relatively low. Towards the borders, somewhat larger deviations are encountered due to singularities and zeros in the relation (\ref{eq:r}). We also checked explicitly that the Hessian of the action has only positive eigenvalues. Finally, notice that the solution we found is in the thin-wall regime, which typically requires a special effort in the shooting algorithm due to a fine-tuning in the release point. As happened in the single-field case \cite{E}, this is 
not an issue with the new action and in fact the sensitivity of the action to changes of the release point is rather small. 

In other cases the escape point might be not so close to the true minimum and finding it is more complicated in the multi-field case as one cannot use the undershooting/overshooting technique. To illustrate such cases we use the same model A with the same choice of parameters except for $\lambda_b=0.07$ (Model A2). Our algorithm finds the escape point without difficulty also in such case and the tunneling path, potential(s) along the path, and EoM crosschecks are shown in Figure~\ref{fig:A2}. In this particular case we get $S\simeq 801$.

For our second model we consider
\be
V(\phi,s) = \lambda (\phi^2 + s^2 - v^2)^2 + \lambda_b v^2 s^2 - \mu^2 \, v \, \phi \, ,\quad (\rm{model \, B}).
\ee
and choose $\lambda=0.2$, $\lambda_b = 0.1$ and $\mu^2 = 0.05 \, v^2$. This model B is not as well motivated as model A
but it demonstrates quite well how the algorithm copes with non-trivial paths in parameter space.
Fig.~\ref{fig:B} shows for this model the tunneling trajectory, potential(s) along it and crosscheck of the solutions
as we did before for model A. The action corresponding to this tunneling trajectory is now $S \simeq 39211$.

\section{Conclusions and Outlook\label{sec:out}} 

The standard calculation of tunneling actions \cite{Coleman} proceeds by finding the $O(4)$ symmetric bounce $\phi_B(r)$, a solution of the Euclidean EoM that interpolates between the false vacuum and the stable phase of the potential and computing its Euclidean action $S_E[\phi_B]$. The Euclidean action has a saddle point at the bounce, with the second functional derivative having one negative mode related to rescalings of the radial coordinate of the bounce. While the one field case can be solved by 'shooting', this is not a viable option in the case of several scalar fields and different methods have been  tried over the years to deal with the problem of finding the saddle point of the multi-field Euclidean action.

An alternative formulation of the calculation of tunneling actions was presented recently in \cite{E}. The new approach does not use 
Euclidean bounces but rather a tunneling potential $V_t$ 
and expresses the action, $S[V_t]$, as a simple integral in field space. In the one field case, it was proven in~\cite{E} that  {\it minimization} of this new action reproduces the standard Euclidean result, with $S[V_t]=S_E[\phi_B]$, effectively getting  rid of the negative rescaling mode of the Euclidean approach. 

Motivated by such interesting behavior, in this paper we have generalized the new formulation to the case of several scalar fields.
We have developed an algorithm for the minimization of the new action and have shown with several numerical examples that this
new approach can be used to calculate such actions in an efficient way. It would be interesting to incorporate such novel approach in 
public tools like VEVACIOUS \cite{VEVACIOUS}.

Besides such direct application, the new approach might also be of 
use to attack other questions of interest, like the large-$N$ scaling
limit of the tunneling action (relevant for discussions of the
string-theory landscape \cite{LargeN1,LargeN2,LargeN3}) and can certainly be modified to the study of tunneling at finite temperature,
simply extending the $d=3$ formulas of \cite{E} to the multi-field case.

\bigskip

\section*{Acknowledgments\label{sec:ack}} 

This work has been supported by the ERC
grant 669668 -- NEO-NAT -- ERC-AdG-2014, the Spanish Ministry MINECO under grants  2016-78022-P and
FPA2014-55613-P, the Severo Ochoa excellence program of MINECO grant SEV-2016-0588 and by the Generalitat de Catalunya grant 2017-SGR-01069.

\section*{Appendix}

In general, the tunneling trajectory is not straight in field space. At each point of the curve ${\bma \phi}(\varphi)$ one can introduce the Frenet-Serret basis of orthonormal vectors
(see {\it e.g.} \cite{Laugwitz} for the tridimensional case and \cite{wikipedia} for its generalization to $N>3$):
\bea
{\bma v_1}&\equiv& \frac{d{\bma \phi}}{d\varphi}\ ,\nonumber\\
{\bma v_2}&\equiv&\frac{1}{\kappa(\varphi)}\frac{d^2{\bma \phi}}{d\varphi^2}\ ,
\eea
where  $\varphi$ is the canonical field along the path, with $|d\vecphi/d\varphi|^2=1$, and
\be
\kappa(\varphi)\equiv \left|\frac{d^2{\bma \phi}}{d\varphi^2}\right|\ ,
\ee
is the curvature of the path. At a given point, the vector ${\bma v_1}$ is tangent to the curve and ${\bma v_2}$ points to the center of curvature of the path at that point. Orthogonality of ${\bma v_1}$ and ${\bma v_2}$ follows from ${\bma v_2}\propto d{\bma v_1}/d\varphi$ and $d ({\bma v_1}\cdot {\bma v_1})/d\varphi=0$.

The rest of unit-vectors are defined by
\be
\overline{\bma v}_m=\frac{d^m{\bma \phi}}{d\varphi^m}-\sum_{n=1}^{m-1}\left(\frac{d^m{\bma \phi}}{d\varphi^m}\cdot {\bma v}_n\right){\bma v}_n\ ,
\ee
and
\be
{\bma v}_m = \frac{\overline{\bma v}_m}{|\overline{\bma v}_m|}\ ,
\ee
and complete and orthonormal basis (being orthogonal as they are constructed by the Gram-Schmidt procedure).

The (generalized) Frenet-Serret formulas give the rate of change of ${\bma v_n}$ when moving along the path:
\be
\left[
\begin{array}{c}
d{\bma v}_1/d\varphi\\
d{\bma v}_2/d\varphi\\
\vdots\\
d{\bma v}_r/d\varphi\\
\vdots\\
d{\bma v}_N/d\varphi
\end{array}
\right]=
\left[
\begin{array}{cccccc}
0 & \chi_1(\varphi) &  &  & & \\
-\chi_1(\varphi) & 0 & \chi_2(\varphi) &  & & \\
 & \ddots &  0&  \ddots & & \\
 & &  -\chi_{r-1}(\varphi)&0  &\chi_{r} (\varphi)& \\
 &  &  & \ddots & 0& \ddots\\
 &  &  &  &  -\chi_{N-1}(\varphi)& 0 
\end{array}
\right]
\left[
\begin{array}{c}
{\bma v}_1\\
{\bma v}_2\\
\vdots\\
{\bma v}_r\\
\vdots\\
{\bma v}_N
\end{array}
\right]\, ,
\ee
with higher order $\chi_i(\varphi)$'s describing higher order derivatives of the curve beyond the first curvature $\kappa(\varphi)$ terms. In the three-dimensional case, for instance, $\chi_2(\varphi)$ defines the torsion, measuring how the curve deviates from the plane orthogonal to ${\bma v_3}={\bma v}_1\times{\bma v}_2$ (planar curves have zero torsion).

Using the Frenet-Serret basis we have
\be
\dot{\bma \phi}=\dot\varphi\ {\bma v_1}\ , \quad 
\ddot{\bma \phi}=\ddot\varphi\ {\bma v_1}+\kappa(\varphi)\ \dot\varphi^2\ {\bma v_2}\ .
\ee
Equation (\ref{VtEoMT}), rewritten as
\be
2\kappa(\varphi)(V-V_t){\bma v_2} = \bma\nabla_\perp V\ ,
\label{VtEoMT2}
\ee
shows that the curvature vector $\bma v_2$ is aligned with $ \bma\nabla_\perp V$, the projection of the potential gradient 
orthogonal to the path.

\end{document}